\documentclass[12pt]{article}
\usepackage{amssymb,amsmath}

\let\ox=\otimes
\def\underdots#1{\mathop{\vtop{\ialign{##\crcr$\hfil\displaystyle{#1}\hfil$\crcr
\dotfill\crcr}}}\limits_{\mathstrut}}
\def\A{\mathbin\land}
\let\iso=\cong
\let\niso=\ncong
\let\MapsIn=\hookrightarrow
\def\oxm{\mathbin{\stackrel{_\mathrm{mat}}{\ox}}}

\newcommand\C{\mathbb{C}}
\newcommand\R{\mathrm{I\!R}}
\renewcommand\H{\mathbb{H}}
\newcommand\Cl{\mathbb{C}\mathrm{l\!l}}
\newcommand{\mbi}[1]{\boldsymbol{#1}}
\newcommand{\mx}[6]{{\left#1\begin{array}{rr}#2&#3\\#4&#5\end{array}\right#6}}
\newcommand{\smx}[6]{{\left#1\begin{smallmatrix}#2&#3\\#4&#5\end{smallmatrix}\right#6}}
\newcommand{\Mx}[3]{\left#1\begin{array}{rrrr}#2\end{array}\right#3}

\newcommand{\qe}[1]{{\mathfrak{\ifcase#1 1\or i\or j\or k\else e_{#1}\fi}}}
\newcommand{\me}[1]{{\mbi{\ifcase#1 1\or u\or v\or w\fi}}}
\newcommand{\e}{{\mbi{e}}}
\renewcommand{\a}{\mbi{a}}
\newcommand{\sgm}{{\mbi\sigma}}
\newcommand{\Id}{{\qe0}}
\newcommand{\Ib}{\mathbf{\dot{\Id}}}
\newcommand{\I}{{\qe1}}
\newcommand{\J}{{\qe2}}
\newcommand{\K}{{\qe3}}
\newcommand{\U}{{\me1}}
\newcommand{\V}{{\me2}}
\newcommand{\W}{{\me3}}
\newcommand{\X}{\sgm_{\!x}}
\newcommand{\Y}{\sgm_{\!y}}
\newcommand{\Z}{\sgm_{\!z}}
\newcommand{\ep}{{\stackrel{_+}{\e}}{}}
\newcommand{\eqdef}{\stackrel{\text{def}}=}
\newcommand{\Mat}[1]{\mathbf{#1}}

\newcommand{\mC}[1]{\C(#1{\times}#1)}
\newcommand{\mR}[1]{\R(#1{\times}#1)}
\newcommand{\EQ}[1]{\mbox{$#1$}}

\newcommand{\ket}[1]{| #1 \rangle}
\newcommand{\bra}[1]{\langle #1 |}
\newcommand{\ktbr}[2]{\ket{#1}\bra{#2}}

\newcommand\0{\ket{0}}
\newcommand\1{\ket{1}}

\begin{document}
%\Large

\title{Quantum Gates and Clifford Algebras.}

\author{{\em Alexander Yu.\ Vlasov}\thanks{%
{\tt E-mail: Alexander.Vlasov@pobox.spbu.ru}}}
\date{12--23 July, 1999}

\maketitle

\begin{abstract}
%\large
 Clifford algebras are used for definition of spinors. Because of using
spin-1/2 systems as an adequate model of quantum bit, a relation of the
algebras with quantum information science has physical reasons. But there
are simple mathematical properties of the algebras those also justifies such
applications.

 First, any complex Clifford algebra with $2n$ generators, $\Cl(2n,\C)$,
has representation as algebra of all $2^n\times 2^n$ complex matrices and so
includes unitary matrix of any quantum $n$-gate. An arbitrary element of
whole algebra corresponds to general form of linear complex transformation.
The last property is also useful because linear operators are not necessary
should be unitary if they used for description of restriction of some unitary
operator to subspace.

 The second advantage is simple algebraic structure of $\Cl(2n)$ that can be
expressed via tenzor product of standard ``building units'' and similar with
behavior of composite quantum systems. The compact notation with $2n$
generators also can be used in software for modeling of simple quantum
circuits by modern conventional computers.

 The standard blocks may be based on three classical groups: $2 \times 2$
complex and real unimodular matrices and group of Weyl spinors, $SU(2)$. The
last group may have more close relation with nonrelativistic quantum systems
as spinor representation of group of 3D rotations, $SO(3)$. The second one,
$SL(2,\R)$, is widely used for theory of quantum error correction codes
\cite{EO} together with a complexification, the first group $SL(2, \C)$ of
Pauli spinors. More exactly, they are based on {\em group algebras}:
$\mC{2}$, $\mR{2}$ and $\H$, {\em i.e.} all $2 \times 2$ complex and real
matrices and quaternions respectively.

\end{abstract}

\newpage

%\section{Introduction}

\section{Clifford algebras.}

\subsection{Preliminaries.}

Generally, Clifford algebra is defined \cite{ClDir} for linear vector space
$V$ with arbitrary quadratic form $Q(\mbi x)$ as some algebra $A$ with map
\EQ{\alpha: V \to A}, \EQ{\alpha(\mbi x)^2 = -Q(\mbi x)\Id} (here $\Id$ is
{\em unit} of the algebra), but let us consider first Euclidean case with
\EQ{Q(\mbi x)=x_1^2+\dots+x_n^2}. The Clifford algebra, \EQ{\Cl(n)},
is generated by $n$ elements \EQ{\e_1,\dots,\e_n} with property:
\begin{equation}
 \e_i \e_j + \e_j \e_i = -2\delta_{ij}; \implies
 \e_i \e_j = - \e_j \e_i \ (i \neq j),\ \  \e_k^2 = -\Id
 \label{ClEu}
\end{equation}
The eq.(\ref{ClEu}) defines $2^n$-dimensional real algebra. It is clear,
because product of any number of $\e_i$ can be
simplified to product\footnote{Let the product is $\Id$ for $k=0$}
with up to $n$ different $\e_i$:
\begin{equation}
 \e_{i_1}\e_{i_2} \cdots \e_{i_k};\quad 0 \leq k \leq n;
 \quad i_1 < i_2 < \dots < i_k
 \label{compos}
\end{equation}
There are $2^n$ such terms because every combination corresponds to
$n$-digits binary number with units in positions \EQ{i_1,i_2,\dots,i_k} :
\[
\underbrace{\underdots{\underbrace{\underbrace{%
00\dots01}_{i_1}00\dots01\vrule}_{i_2}0\dots0}1\vrule}_{i_k}00\dots00
\]

Elements of vector space $V$ map to $n$-dimensional subspace of the algebra:
\EQ{\mbi{v} = \sum_{i=1}^n v_i \e_i}. Elements with unit norm \EQ{\|\mbi v\|=1}
\footnote{More generally \EQ{|Q(\mbi v)|=1}} have inverse, (for \EQ{\Cl(n)},
\EQ{\mbi v^{-1} = -\mbi v} because \EQ{-\mbi v{\cdot}\mbi v = \Id}) and product of
even number of such elements is some group. It is called {\em spinor group}
\EQ{Spin(n)} and it has 2-1 isomorphism with group of
$n$-dimensional rotations, $SO(n)$.

\smallskip

In the definition of spinor group is used only multiplication of {\em even}
number of elements of algebra \EQ{\Cl(n)}. It is possible to build $2^{n-1}$
different compositions eq.(\ref{compos}) with even number of generators of
\EQ{\Cl(n)} and all such compositions produce {\em even subalgebra} \EQ{\Cl^e(n)}
of Clifford algebra and really spinor group is subspace of the smaller algebra.
But it can be shown that \EQ{\Cl^e(n)} is isomorphic with \EQ{\Cl(n-1)} and the
algebra also may be used for representation of \EQ{Spin(n)} group. The property
is mentioned here to explain why it is important sometime to consider Clifford
algebras with dimension reduced by unit, for example \EQ{\Cl(2)} is enough to
build spinor group for 3D rotations.

\medskip

It is possible also to introduce yet another algebra, \EQ{\Cl_+(n)}, if to
change sign of square of all $\e_k$ to $+\Id$ :
\begin{equation}
 \e_i \e_j + \e_j \e_i = 2\delta_{ij}; \implies
 \e_i \e_j = - \e_j \e_i \ (i \neq j),\ \  \underline{\e_k^2 = \Id}
\end{equation}
The two real algebras are not equivalent.

Clifford algebra \EQ{\Cl(m,l)} corresponds to more general, pseudo-Euclidean
case (\EQ{{\e_k}^2 = \pm \Id}):
\begin{equation}
 \e_i \e_j = - \e_j \e_i \ (i \neq j), \quad
 \e_k^2 = -\Id \ (k \leq m), \quad
 \e_k^2 = \Id \ (k > m)
\end{equation}
Two previous examples are special cases of the definition:
\EQ{\Cl(n) = \Cl(n,0)}, \EQ{\Cl_+(n) = \Cl(0,n)}.

It is possible also to consider `degenerated' case ( \EQ{Q(x) \equiv 0 } ) :
\begin{equation}
 \e_i \e_j + \e_j \e_i = 0 \quad \forall i,j \ (\e_k^2=0)
\end{equation}
It is known also as {\em Grassmann algebra} $\Lambda^n$ (with notation
\EQ{a \A b} is used instead of \EQ{a\,b}, \EQ{\e_i\A\e_j = -\e_j\A\e_i}) and usually
it is considered separately. One of applications of Grassmann algebras is
related with description of subspaces of $n$-dimensional vector space.

\medskip

It was real Clifford algebras and the complexification of any algebra
\EQ{\Cl(l,n-l)} produces the same universal Clifford algebra \EQ{\Cl(n,\C)}
because it is always possible to `correct' sign of any $\e_k^2$ by
substitution \EQ{\tilde \e_k = i \e_k}, \EQ{\tilde\e_k^2 = -\e_k^2} .

Let us use notation $\e_k$ for elements of basis with \EQ{\e_k^2 = -\Id} and
$\ep_k$ for \EQ{\ep_k^2 = \Id}, \EQ{\ep_k = i\e_k}.

%\medskip
\noindent ---

 The following constructions of Grassmann algebras are used further:
Let us consider even-dimensional
universal Clifford algebra \EQ{\Cl(2n,\C)}. It is possible to build a
Grassmann algebra $\Lambda^n$ with using $n$-generators:
\begin{equation}
 \mbi d_l = \e_{2l} + i \e_{2l+1}; \implies
 \mbi d_k \mbi d_j = -\mbi d_j \mbi d_k \quad \forall k,j \quad (\mbi d_k^2=0)
\end{equation}

It is possible also to define two different algebras: $\Lambda^n_{_+}$
with generators~$a_l$ and $\Lambda^n_{_-}$ with $a_l^\dag$:
\begin{equation}
 \a_l = \frac{\ep_{2l} + i \ep_{2l+1}}{2} \,, \quad
 \a^\dag_l = \frac{\ep_{2l} - i \ep_{2l+1}}{2}
\label{aadag}
\end{equation}
with properties:
\begin{equation}
 \fbox{$\quad%
 \{ \a_i, \a_j \} = 0,\quad \{ \a^\dag_i, \a^\dag_j \} = 0, \quad
 \{ \a^\dag_i, \a_j \} = \delta_{ij}
 \quad$}
\label{aacomm}
\end{equation}
where \EQ{\{a,b\} \equiv (a\,b + b\,a)}.
A nontrivial case is
\EQ{\{\a^\dag_l,\a_l\} = \Id} also can be simply checked:
\begin{align*}
  4\,\a_l\,\a^\dag_l & = (\ep_{2l} + i \ep_{2l+1})(\ep_{2l} - i \ep_{2l+1}) = \\
& = \ep_{2l}^2 + i \ep_{2l+1}\ep_{2l} - i \ep_{2l}\ep_{2l+1} + \ep_{2l+1}^2
 = 2 - 2i\,\ep_{2l}\ep_{2l+1} \\
  4\,\a^\dag_l\,\a_l & = (\ep_{2l} - i \ep_{2l+1})(\ep_{2l} + i \ep_{2l+1}) = \\
&= \ep_{2l}^2 - i \ep_{2l+1}\ep_{2l} + i \ep_{2l}\ep_{2l+1} + \ep_{2l+1}^2
 = 2 + 2i\,\ep_{2l}\ep_{2l+1}
%\\ &\qquad 4 \a_l\,\a^\dag_l + 4 \a^\dag_l\,\a_l = 4
\end{align*}

\subsection{Two-dimensional case.}

Let us consider real Clifford algebras first. The 4D real algebra \EQ{\Cl(2)} is
described by elements \EQ{\I = \e_1}, \EQ{\J = \e_2},
\EQ{\K = \e_1 \e_2} and unit element $\Id$. The elements satisfy relations:
\EQ{\I^2 \eqdef -\Id}, \EQ{\J^2 \eqdef -\Id}, \EQ{-\J\I \eqdef \I\J \eqdef \K},
\EQ{\K^2 = \I\J\K = -\Id}, \EQ{\I\K = \I\I\J = -\J = \J\I\I = -\K\I},
\EQ{\J\K = -\J\J\I = \I = -\I\J\J = -\K\J} and the equations define well known
4D {\em quaternions algebra} $\H$
%,\EQ{\mathbf q = q_0\Id+q_1\I+q_2\J+q_3\K \in \H}
introduced by Hamilton at 1843 as a generalization of complex numbers.

\smallskip

Another 4D real Clifford algebra \EQ{\Cl_+(2)} is generated by elements $\U$,
$\V$, \EQ{\W = \U\V} with defining relations are:  \EQ{\U^2 \eqdef \Id},
\EQ{\V^2 \eqdef \Id}, \EQ{\U\V \eqdef -\V\U \eqdef \W} and with implications:
\EQ{\W^2 = \U\V\W = -\Id}, $\V\W = \V\V\U = \U = \U\V\V = -\W\V$,
$\U\W = -\U\U\V = -\V = -\V\U\U= -\W\U$.

It is possible to consider $\V$ and $\W$ as generators of some new Clifford
algebra and because \EQ{\W^2 = -\Id} it is \EQ{\Cl(1,1)} and so the algebra is
isomorphic with \EQ{\Cl_+(2) \equiv \Cl(0,2)}.

The \EQ{\Cl_+(2)} is isomorphic with 4D algebra \EQ{\mR{2}} of all \EQ{2\times2} real
matrices. It is enough to choose: \EQ{\U = \smx (0110)}, \EQ{\V = \smx (100{-1})},
\EQ{\W = \U\V =-\V\U = \smx (0{-1}10)} and \EQ{\Id = \smx (1001)} is unit matrix.

\medskip

The complexification of the algebras, \EQ{\Cl(2,\C)} is isomorphic with
algebra \EQ{\mC{2}} of all complex \EQ{2\times2} matrices and representation
by Pauli matrices is: \EQ{\ep_1 = \X}, \EQ{\ep_2 = \Y}, \EQ{\ep_1\ep_2 = i\Z},
\begin{equation}
 \X = \mx(0110), \quad \Y = \mx(0{-i}i0), \quad \Z = \mx(100{-1}).
\end{equation}
with standard relations \EQ{\X^2=\Y^2=\Z^2=\Id}, \EQ{\X\Y=i\Z=-\Y\X},
$\Z\X = i\Y = -\X\Z$, \EQ{\Y\Z = i\X = -\Z\Y}.

\smallskip

The real Clifford algebras \EQ{\Cl(2) \equiv \Cl(2,0) \iso \H} and
$\Cl_+(2) \equiv \Cl(0,2) \iso \Cl(1,1) \iso \mR{2}$ are subalgebras
of \EQ{\Cl(2,\C) \iso \mC{2}}:
\begin{align}
 \Cl_+(2) \MapsIn \Cl(2,\C):&\quad \U \to \X,\ \V \to \Z,\ \W \to \Y/i
 \label{RPa}\\
 \Cl(2) \MapsIn \Cl(2,\C):&\quad \I \to i\X,\ \J \to i\Y,\ \K \to i\Z
 \label{quPa}
\end{align}
Really, the eq.(\ref{quPa}) corresponds to usual representation of quaternions
by complex \EQ{2\times2} matrices: \EQ{\I = \smx (0ii0)}, \EQ{\J = \smx (01{-1}0)},
\EQ{\K = \smx (i00{-i})}, but it is useful sometime to do not use some
particular matrix notation and consider quaternions as an {\em abstract}
linear algebra defined by relation with $\I$, $\J$ and $\K$ introduced
earlier:
\begin{equation}
  \I^2 = \J^2 = -\Id,\ \I\cdot\J=-\J\cdot\I = \K;
  \ q_0\Id+q_1\I+q_2\J+q_3\K \in \H;\ (q_k \in \R)
\end{equation}
For example, such definition makes more clear difference between 4D {\em real}
algebra of quaternions and 4D complex Pauli algebra.

It should be mentioned, that the Pauli algebra can be considered also as 8D
{\em real} Clifford algebra \EQ{\Cl_+(3)} with relations: \EQ{\X^2=\Y^2=\Z^2=\Id}.
An element \EQ{\Ib \equiv \X\Y\Z} of the algebra commutes with any
other element and corresponds to \EQ{i\Id} of complex algebra \EQ{\Cl(2,\C)}.

\subsection{Constriction of \EQ{\Cl(2n,\C)}.}

Let us recall that for $n$-dimensional linear space $V$ with basis
\EQ{e_1,\dots,e_n} and $m$-dimensional linear space $W$ with basis
\EQ{e'_1,\dots,e'_m} the {\em tensor product} \EQ{V \ox W} of the spaces
can be introduced as \EQ{m{\cdot}n}-dimensional space with basis: \EQ{e_i\ox e'_j}.

Similarly, for $n$-dimensional linear space $V$ with basis \EQ{e_1,\dots,e_n}
the {\it tensor power} \EQ{V^{{\ox}k}} of $k$ copies of such space
can be introduced as $n^k$-dimensional space with basis:
\EQ{e_{i_1}\ox e_{i_2}\ox\dots\ox e_{i_k}}, $1 \leq i_l \leq n$.

If the linear spaces have structure of some algebras $A$ and $B$ it is
possible to build new \EQ{nm}-dimensional algebra \EQ{A \ox B} with composition is
defined as: \EQ{(a \ox b) \cdot (c \ox d) \equiv (ac) \ox (bd)}
for elements of basis and extended to all elements by distributivity. The
$n^k$ dimensional tensor power \EQ{A^{{\ox}k}} is defined analogously.

\medskip

Let us show direct construction of \EQ{\Cl(2n,\C)} as algebra \EQ{\mC{2^n}} of all
\EQ{2^n\times2^n} matrices. It can be done by two steps with using Pauli
matrices. First step is isomorphism \EQ{\Cl(2n,\C) \iso \mC{2}^{{\ox}n}} and
second is \EQ{\mC{2}^{{\ox}n} \iso \mC{2^n}}.

It is possible to show first isomorphism by direct construction of \EQ{2n}
generators of \EQ{\Cl(2n,\C)} :
\begin{equation}
\begin{split}
 \ep_{2k} = &
 \underbrace{\Id\ox\dots\ox\Id}_{n-k-1}\ox\X\ox\underbrace{\Z\ox\dots\ox\Z}_k
 \\ \ep_{2k+1} = &
 \underbrace{\Id\ox\dots\ox\Id}_{n-k-1}\ox\Y\ox\underbrace{\Z\ox\dots\ox\Z}_k
\end{split}
\label{genClnnC}
\end{equation}
To check that all the \EQ{2n} generators are anticommutative it is possible to
introduce the construction inductively. For \EQ{n=1} the \EQ{\Cl(2,\C)} has two
generators $\X$ and $\Y$. If the suggestion is shown for some \EQ{l \geq 1}
and we have \EQ{2l} generators $\ep_1,\dots,\ep_{2l} \in \mC{2}^{{\ox}l} \iso
\Cl(2l,\C)$ then for \EQ{l+1} it is also possible to choose \EQ{2l+2} generators
in \EQ{\mC{2}^{{\ox}l+1}}: {\em i.e.} \EQ{2l} anticommutative elements
\EQ{\ep_1\ox\Z,\dots,\ep_{2l}\ox\Z} together with two elements
\EQ{\Id^{{\ox}l}\ox\X} and \EQ{\Id^{{\ox}l}\ox\Y}, where
\EQ{\Id^{{\ox}l} = {\underbrace{\Id\ox\dots\ox\Id}_l}}
%!!!! smash
is unit of \EQ{\Cl(2l,\C)}. So \EQ{\Cl(2l+2,\C) \iso \mC{2}^{{\ox}l+1}}.

\medskip

The isomorphism of \EQ{\mC{2}^{{\ox}n} \iso \mC{2^n}} can be also shown by
induction. It is true for \EQ{n=1} and if it is proven for some \EQ{l \geq 1}
then any \EQ{A \in \mC{2}^l \iso \mC{2^l}} is represented by some
\EQ{2^l\times2^l} matrix \EQ{\Mat A}. It is possible to check that
\EQ{\mC{2^{l+1}} \iso \mC{2} \ox \mC{2^l}}, if to define operation:
\begin{equation}
\begin{split}
 &\Mat a = \mx({a_{11}}{a_{12}}{a_{21}}{a_{22}}) \in \mC{2},
  \qquad \Mat A \in \mC{2^l},  \\
 &\Mat a \oxm \Mat A \equiv
  \mx\lgroup{(a_{11}\Mat A)}{(a_{12}\Mat A)}
            {(a_{21}\Mat A)}{(a_{22}\Mat A)}\rgroup  \in \mC{2^{l+1}}
\end{split}
\label{MatOx}
\end{equation}
It can be checked directly,
\EQ{(\Mat{a\oxm A}) \cdot (\Mat{b\oxm B}) = \Mat{ab\oxm AB}} if $\Mat a,\Mat b$
are basis elements: $\Id,\X,\Y,\Z$ and it is extended for all elements by
distributivity.

\section{Quantum gates.}

Any quantum $n$-gate is described by some unitary \EQ{2^n\times2^n}
matrix. It acts on $n$-qubits in $2^n$-dimensional complex vector space.

Matrices of all \EQ{2^{2n}=4^n} possible compositions of \EQ{2n} generators in
eq.(\ref{genClnnC}) are unitary. Really, any such element has form:
\begin{equation}
 \e_{l_1}\e_{l_2}\dotsm\e_{l_k} \xrightarrow{0 \leqslant k \leqslant 2n}
 \e_{\mbi I} \equiv \sgm_{\!i_1}\ox\sgm_{\!i_2}\ox\dots\ox\sgm_{\!i_n}
\label{composC}
\end{equation}
where \EQ{\sgm_{\!i_k}} is one of four basis elements: \EQ{\Id,\X,\Y,\Z}. In matrix
representation used here \EQ{\sgm_i^* = \sgm_i}, \EQ{\sgm_i^2 = \Id}. For any
\EQ{\sgm_i} and matrix \EQ{\Mat A^* = \Mat A} the eq.(\ref{MatOx}) implies
\EQ{(\sgm_i\oxm\Mat A)^* = (\sgm_i\oxm\Mat A)} and so
\EQ{\e_{\mbi I}^* = \e_{\mbi I}}. Because also \EQ{\e_{\mbi I}^2 = \Id}, all
the matrices for \EQ{4^n} elements satisfy equations:
\EQ{\e_{\mbi I}^* = \e_{\mbi I}= \e_{\mbi I}^{-1}}, they
are not only Hermitian (\EQ{\Mat A^* = \Mat A}), but unitary too
(\EQ{\Mat A^* = \Mat A^{-1}}).

Such construction of unitary matrices may be not very general, because
it produces finite number of gates, but it is possible to consider some
analogous of `Hamiltonian approach', if to use Hermitian matrices
$\Mat H$ and then \EQ{\Mat U = \exp (i\,\Mat H)} is unitary. It is useful
because whole \EQ{4^n} dimensional space of Hermitian matrices is
produced by combinations up to $4^n$ elements $\e_{\mbi I}$ described
by eq.(\ref{composC}) with real coefficients:
\begin{equation}
 \mathcal H = \left\{ \Mat H \ : \ \Mat H =
 \sum\nolimits_{\mbi I}{c_{\mbi I} \e_{\mbi I}};
 \quad c_{\mbi I} \in \R \right\}
\end{equation}
\EQ{\Mat H \in \mathcal H}, \EQ{\dim \mathcal H = 4^n}.

\medskip

It is necessary also to define action of an element of $\Cl(2n,\C)\iso\mC{2^n}$
on $n$-qubit register as an action of \EQ{2^n\times2^n} matrix on complex
vector in $2^n$-dimensional complex space. Let us define it for basis
of $2^n$ vector space
\begin{equation}
 e_L = e_{l_1}\ox e_{l_2}\ox\dots\ox e_{l_n},\quad l_k = 0,1
\end{equation}
or with notation used in quantum information science
\begin{gather}
 \ket{L} = \ket{l_1\,l_2 \dots l_n} =
 \ket{l_1}\ox \ket{l_2}\ox\dots\ox \ket{l_n},\quad l_k = 0,1
 \label{nqubasis}\\
 e_0=\ket{0} = \Mx({1\\0}),\quad e_1=\ket{1} = \Mx({0\\1})
 \label{qubasis}
\end{gather}
and $4^n$-dimensional basis eq.(\ref{composC}) of
Clifford algebra:
\begin{equation}
 \mbi{g}_I = \sgm_{\!i_1}\ox \sgm_{\!i_2}\ox\dots\ox \sgm_{\!i_n},
 \quad \sgm_{\!i_k} \in \Id,\X,\Y,\Z
\end{equation}
The action is defined for the elements of basis component-wise:
%\begin{equation}
% \mbi{g_I} (e_L) = (\sgm_{\!i_1} e_{l_1}) \ox (\sgm_{\!i_2} e_{l_2})\ox
% \dots\ox (\sgm_{\!i_n} e_{l_n})
% \label{eIeL}
%\end{equation}
\begin{equation}
 \mbi{g}_I \ket{L} = \bigl(\sgm_{\!i_1} \ket{l_1}\bigr) \ox
  \bigl(\sgm_{\!i_2} \ket{l_2}\bigr)\ox
  \dots\ox \bigl(\sgm_{\!i_n} \ket{l_n}\bigr)
 \label{eIeL}
\end{equation}
where \EQ{\sgm_{\!i_k} \ket{l_k}} is action of \EQ{2\times2} complex
matrix $\sgm$ on 2D complex vector $\0$ or $\1$. The action of arbitrary
element (gate) to a general vector ({\em entangled} state) is produced from
eq.(\ref{eIeL}) by linearity:
\begin{equation}
 \mbi G \ket{\psi} =
 \sum_I{a_I\mbi{g}_I}\, \sum_L{c_L\ket{L}} =
 \sum_{I,L}{(a_I c_L)\,\mbi{g}_I\ket{L}}
\end{equation}
\medskip

 So, any linear transformation of $2^n$-dimensional space can be represented
by elements of Clifford algebra with $2n$ generators
\begin{equation}
 \Cl(2n,\C) \iso \mC{2^n} : \C^{2^n} \to \C^{2^n}
\end{equation}
To build particular matrix for some quantum $n$-gate it is possible to
start with simple $2\times2$ $\sgm$-matrices for $\Cl(2,\C)$ and qubit
in basis eq.(\ref{qubasis}). After it, recursively $1,\dots,n$ on each
step with using eq.(\ref{MatOx}) it is considered matrices and vectors
with doubled size:
\begin{equation}
  \bigl(\square\bigr)\cdot\ket{...} \xrightarrow{k \to k+1}
  \mx\lgroup{(\square)}{(\square)}
            {(\square)}{(\square)}\rgroup \!\cdot
  \Mx|{\!\bigl(\ket{0...}\bigr)\!\! \\ \!\bigl(\ket{1...}\bigr)\!\!}>
\end{equation}

\medskip

For example of application the theory of Clifford algebras to quantum
information science let us use now other basis with dual Grassmann
elements introduced by eq.(\ref{aadag},\ref{aacomm}) to describe approach
with occupation numbers used by Feynman \cite{feynman:simul,feynman:comp}.

Let us apply eq.(\ref{aadag}) to expressions for Clifford generators
eq.(\ref{genClnnC}):
\begin{equation}
\begin{split}
 \a_k = & \Id^{{\ox}n-k-1}\ox\frac{\X+i\Y}2\ox\Z^{{\ox}k} \\
 \a^\dag_k = & \Id^{{\ox}n-k-1}\ox\frac{\X-i\Y}2\ox\Z^{{\ox}k}
\end{split}
\label{GenFeym}
\end{equation}

The matrices \EQ{\a = \frac{\X+i\Y}2 = \smx (0100)} and
\EQ{\a^\dag = \frac{\X-i\Y}2 = \smx (0010)} correspond to {\em annihilation}
and {\em creation} operators used by Feynman for description of quantum
bit and commutation relations in eq.(\ref{aacomm}) correspond to correct
extension of the operators for $n$ fermions.

\medskip

The approach is analogue of {\em second quantization} and some
advantages of the paradigm are discussed already in \cite{RQC}.

On the other hand, due to isomorphism \EQ{\Cl(2n,\C) \iso \mC{2^n}} described
above, the same elements produce basis for description of quantum gates by
\EQ{2^n\times2^n} complex matrices used in S-matrix approach originated
by Deutsch \cite{deutsch:gates} {\em et al}.

\smallskip

The operator $\a$ of annihilation and $\a^\dag$ of creation can be also
written as (here {\bf 0} is zero vector):
\begin{align}
 &\a = \ktbr 01; &&\a^\dag = \ktbr 10 \\
 &\a \1 = \0,\quad \a \0 = \mathbf 0; &
 & \a^\dag \0 = \1,\quad \a^\dag \1 = \mathbf 0
\end{align}

\section{Some remarks}

The construction of tensor product of algebras depends on {\em field
of scalars} (say $\R$ or $\C$) used in definition of the algebras. And if
algebras could be considered either as complex or as real, then it
is possible to use two kinds of tensor products:
\begin{equation}
 A \ox_{_\C} B \niso A \ox_{_\R} B
\end{equation}
For example,
for algebra of Pauli matrices considered as 8D real algebra the $n$-th
tensor power is \EQ{8^n = 2^{3n}}D real algebra and it is not equivalent
to construction used in the paper with $2^{2n}$ complex or $2^{2n+1}$
real dimension. The fact clarify difference between spaces used in the
paper and in \cite{Cory} there real tensor product was used.

\medskip

With using of real tensor product it is also possible to express space of
quantum $n$-gates and $\Cl(2n,\C)$ :
\begin{equation}
 \mbi{G}_n \in \C\ox\mR{2}^{{\ox}n}
\end{equation}
or:
\begin{equation}
 \mbi{G}_n \in \C\ox\H^{{\ox}n}
\end{equation}
where $\C$ is considered as 2D real and $\H$ as 4D real algebras.

\section*{Acknowledgements}
Author is grateful to organizers of {\em TMR Network School on
``Quantum Computation and Quantum Information Theory''} (Italy, Torino,
12--23 July 1999) for invitation, financial support and hospitality.


\begin{thebibliography}{99}
 \bibitem{EO} A. R. Calderbank, E. M. Rains, P. W. Shor, N. J. A. Sloane,
  Quantum Error Correction and Orthogonal Geometry, {\em Phys. Rev. Lett.
  \bf 78} (1997), 405--408.
 \bibitem{ClDir} J. E. Gilbert, M. A. M. Murray, {\em Clifford algebras and
  Dirac operators in harmonic analysis}, Cambridge University Press, 1991.
 \bibitem{feynman:simul} Richard Feynman, Simulating Physics with
  Computers, {\em Internal Journal of Theoretical Physics \bf 21} (1982),
  467--488.
 \bibitem{feynman:comp} Richard Feynman, Quantum-Mechanical
  Computers, {\em Foundations of Physics \bf 16} (1986), 507--531.
 \bibitem{deutsch:gates} David Deutsch, Quantum computational networks,
  {\em Proceedings of Royal Society of London \bf A 425}, (1989), 73--90.
 \bibitem{RQC} A. Yu. Vlasov, Quantum theory of computation and relativistic
  physics, {\em Proceedings of the Fourth Workshop on Physics and
  Computation} (1996), 332--333 [quant-ph/9701027 -- full version]
 \bibitem{Cory} S. S. Somaroo, D. G. Cory, T. F. Havel, Expressing the
  operations of quantum computing in multiparticle geometric algebra,
  {\em Phys. Lett. \bf A 240} (1998), 1--7.
\end{thebibliography}
\end{document}